# A brief overview of Türkiye Earthquake: insight into the building damage

Rajib Biswas*

Corresponding author: rajib@tezu.ernet.in

Department of Physics, Tezpur University, Tezpur-784028, Assam

**Abstract**: The month of February, 2023 had been a nightmare for the people of Türkiye. On 6$^{th}$ Feb, there was a devastating earthquake which jolted Türkiye like never before. The aftershock which followed later added more fuel to the overall damages. Till now, there has been reports several aftershocks which is believed to continue for another considerable span of time. In terms of damage; the loss was totally irreparable. Millions of people were rendered homeless and a large chunk of population had lost their lives due to building collapse. This mini communication overviews the past seismicity of Türkiye. As per preliminary report, there has been substantial liquefaction and ground subsidence in and around the epicenters. Accordingly, the liquefaction is also briefly detailed along with types of prevalent sediments Further, the damages as well as building collapsed are detailed here along with probable causes as well as lacunae observed in building construction. Future strategies may possibly involve such as—base isolation or isolators can be introduced in order to make the buildings more resilient earthquakes. However, strict monitoring and compliance of structures to building code should be implemented with strict measures. Any violation of such codes should be penalized. All these measures can lead to earthquake resilient society.

1. Introduction

A 7.8-magnitude earthquake of February 6, 2023, occurred in southern Turkey, close to Syria's northern border. A magnitude 6.7 aftershock occurred 11 minutes after the initial earthquake. An earthquake of magnitude 7.8 was caused by shallow strike-slip faulting. A near-vertical left-lateral fault striking northeast-southwest or a right-lateral fault striking southeast-northwest were both ruptured by the event. According to preliminary information, the earthquake occurred close to a triple-junction of the African, Arabian, and Anatolia plates. The earthquake's mechanism and epicentre are consistent with it having happened on either the Dead Sea transform fault zone or the East Anatolia fault zone. Turkey's westward extrusion into the Aegean Sea is accommodated by the East Anatolia fault, and the Arabian Peninsula's northward motion in relation to the African and Eurasian plates is accommodated by the Dead Sea Transform [1-6].

In 10 provinces across Turkey, there were at least 46,104 fatalities, 114,991 injuries, approximately 1,5 million people made homeless, at least 164,000 buildings severely damaged or destroyed, and 150,000 commercial facilities considerably affected. At least 490 structures were demolished and many more were damaged in northwest Syria, resulting in at least 6,795 deaths, 14,500 injuries, and 5,37 million people being made homeless. Turkey's Golbasi and Hatay experienced liquefaction and land subsidence. With wave heights of 17 cm at Famagusta, Cyprus; 13 cm at Erdemli; and 12 cm at Iskenderun, Turkey, a minor tsunami was produced. IXth highest intensity. Despite the fact that they are frequently represented on maps as single spots, earthquakes actually rupture planes with dimensions. A fault that is 190 km long and 25 km wide is frequently ruptured by a magnitude 7.8 strike slip earthquake. Figure 1 shows the waveform of the main event.

The aim of this short communication is to briefly overview the seismicity of Türkiye in relation to the devastating main event assisted by liquefaction observation along with the outlining of the causes of building damages which resulted in huge fatalities in the decade so far. Accordingly, the first section deals past seismicity. The 2$^{nd}$ section dwells upon liquefaction pattern observed due to earthquake. The third section analyzes the building damage followed by recommendations.

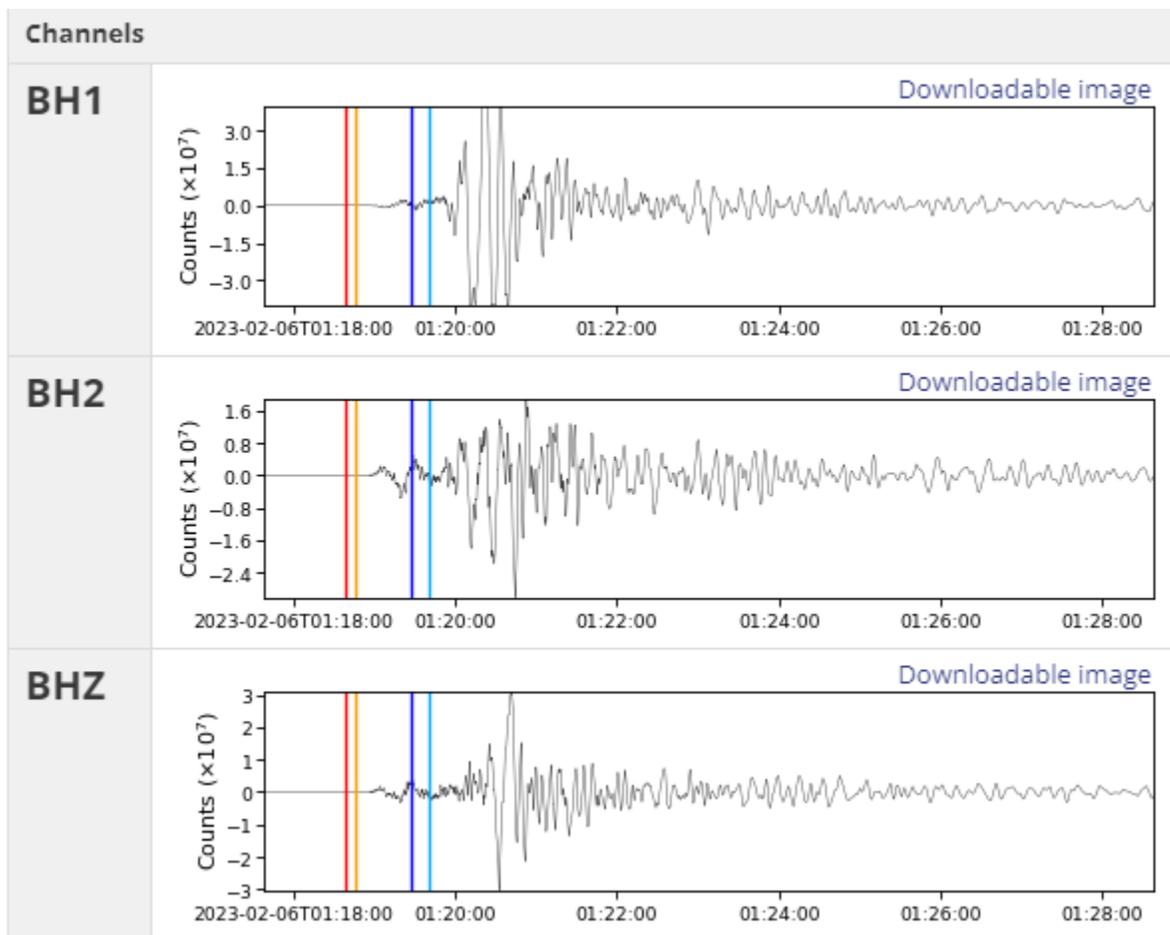

Figure 1. Seismic waveform of the M_W7.8 earthquake (Courtesy: USGS)

## 2. Past seismicity

The earthquake that happened on February 6 occurred in a seismically active area. Since 1970, there have only been three earthquakes with a magnitude of 6 or higher that have happened within 250 kilometers of the February 6 quake. On January 24, 2020, the largest of these, with a magnitude of 6.7, occurred to the northeast of the February 6 earthquake. These earthquakes all happened around or along the East Anatolia fault. Southern Turkey and northern Syria have previously been subjected to big and destructive earthquakes, notwithstanding the relative seismic quiescence of the epicentral region of the February 6 earthquake. Although the particular locations and magnitudes of these earthquakes are unknown, Aleppo, in Syria, has traditionally been devastated by big earthquakes. In 1138 and 1822, respectively, earthquakes with estimated magnitudes of 7.0 and 7.1 both hit Aleppo. There were 20,000–60,000 fatality estimates for the earthquake of 1822 [1-5].

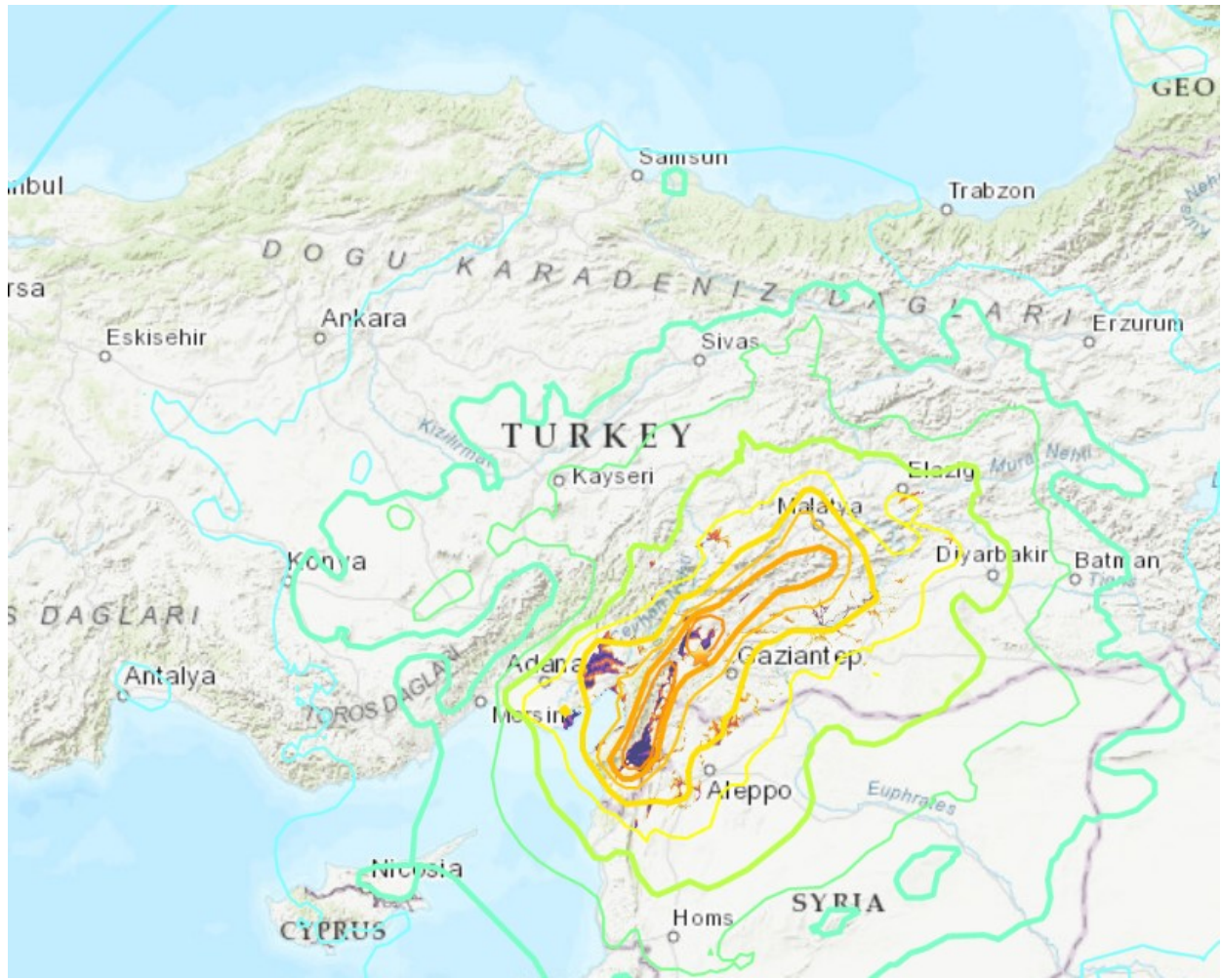

Figure 2. Liquefaction Map (Courtesy: USGS)

Figure 2 depicts the level of liquefaction as observed. Unexpectedly, most liquefaction and lateral spreading sites manifested along and near coastal sites, fluvial valleys and drained lake/swamp areas, covered by Holocene loose sediments. The distribution of mapped liquefaction sites along the fault rupture and shows that most concentrations are found in Holocene sediment-filled basins [4-5].

### 3. Looking into building damages: probable loopholes

The Turkey earthquake was an eye opener. The two earthquakes 7.8 and 7.5 actually devastated the infrastructure of two urban settlements. It was such a massive earthquake that all buildings nearly suffered major to minor damages. To be precise, a copious no. of buildings had given in and collapsed in response to the heavy shaking of the two earthquakes. As the saying goes, earthquakes do not kill people; but buildings do. The statement has been proven again.

Each disaster provides lessons. There is a need to strengthen building, seismic codes. Past earthquakes led Turkey to revamp their code in 1998 as well as 2007. Although in practice, however, reinforcement problem impairs it to a great level.

The capital, Istanbul's booming population and rapid construction are a bottleneck for the ensuing rescue operations.

Design codes and professional codes when implemented by architects, engineers and contractors can clearly reduce risks and thus lives and damages. Structures must be maintained. Poor maintenance which often occurs with infrastructure as a cost saving measure will reduce effective mess of a structural improvement. Similarly, structures have design lives. They wear out over time, especially if poorly maintained. Many disasters are highly foreseeable. Just because, it has not happened before, does not mean safe. Yet, infrastructure is often neglected and overlooked as long as no problems have occurred.

Looking at the pattern of building damages that occurred in Turkey, most of them are soft story building. There were ground level parking with pillars in between them. If we look at the tectonic settings of turkey; it is quite complex. There are three plates which actually collided with each other leading to the occurrence of these huge earthquakes.

As a strict action, the Govt. of Turkey had been on a mission to identify the unscrupulous builders—engaged in the construction o these damaged structures due to earthquake. Accordingly, a considerable chunk of people has been summoned in order to execute criminal proceedings against them. Now the pertinent question is—is this enough? Damage as well as huge causalities have already occurred. What had already happened cannot be undone.

We know that every high-rise structure is prone to shaking. All such structures have a fundamental frequency. When seismic waves travel through a medium of low density; its velocity decreases. Thereby, it spends more time in that medium. As a result, if the predominant frequency matches with frequency of the above structure; there occurs resonance and the building sway with maximum amplitude—leading to eventual collapse.

Buildings that collapsed during the tragedy because of poor construction, inferior materials, and a failure to adhere to building rules have sparked indignation across the country. The earthquake caused some brand-new apartment buildings, which were marketed as being built to the greatest earthquake specifications, to collapse.

Experts are attempting to piece together data on the compliance level of buildings in the geologically vulnerable region more than a month after three fatal earthquakes slammed Turkey. According to a recent report summarizing the first results of the damage assessment following the earthquakes, violations of the building code's requirements during the previous 20 years were a major factor in the significant loss of life and infrastructure damage.

The Turkish Earthquake Code (2018) design levels were not met by the buildings in the provinces of Gaziantep, Hatay, Kahramanmaras, and Adiyaman, according to the report by a team of scientists from Middle East Technical University (METU), Ankara, and colleagues. Following the destruction, Turkey's building codes and construction methods have come under criticism.

Notwithstanding the fact that the earthquakes were exceptional, the experts highlighted that buildings should have survived and not collapsed in the manner that they did.

According to the international team's "Preliminary Reconnaissance Report," buildings built after 2002 are likely to fare better during earthquakes than earlier structures. The analysis reveals that more than 1,000 buildings built after 2000 suffered significant damage or collapsed, contravening the performance aim set forth in the code that evaluates the seismic risk of a building in relation to its location. This looked to be a significant observation, according to the report, necessitating more research into the caliber of those structures' design and construction. Again, inadequacies could also emerge due to the existence of soft stories—which are entrances or basements without continuous walls with those of the upper storeys. It was found that the "pancake" collapses of numerous structures were, in fact, caused by inadequate foundations. This type of structural collapse known as a "pancake collapse" happens when upper floors of a building sink into lower ones. Examples of "severe alterations" that were categorically unacceptable in the amended requirements include the use of low-quality materials, unribbed reinforcement bars used in construction, and insufficient stirrup tightening (which is intended to laterally constrain steel reinforcement). Many new structures constructed after 2000 that were not adequately engineered, well inspected, or whose soil-structure link remained unestablished, were damaged or destroyed beyond expectations. 6 cm-long stones were found in concrete samples recovered from a fallen structure in Adiyaman. These were employed to bulk out the concrete and came from a nearby river.

4. Way Forward

Given the level of damage that had occurred due to the earthquakes, the main agenda before Türkiye Govt. is to rebuild the area which may amount to $100 billion—as per UN reports. The common mass should be more aware and vigilant so that maximum damage can be reduced. The fundamental tenet of construction is to let some degree of damage inside the structure. This damage guarantees that the building still stands erect but does not collapse by absorbing the earthquake's force. Likewise, it is possible to include elements like dampers, which function as shock absorbers

as the building sways, and rubber bearings, which are installed underneath buildings and absorb earthquake energy. Similarly, base isolation or isolators can be introduced in order to make the buildings more resilient earthquakes. However, strict monitoring and compliance of structures to building code should be implemented with strict measures. Any violation of such codes should be penalized.

5. Final Remarks

This brief communication overviews the recent Türkiye Earthquake and liquefaction caused thereof. The past seismicity is also delineated. The probable causes/pitfalls associated with collapsed structures are reviewed. As we know, earthquakes can not be averted, however, the seismic hazard can be mitigated. With strict measures as well as earthquake resilient structures, the devastated nation Türkiye is believed to recover in the coming days.

References:

1. https://www.bbc.com/news/science-environment-64920236
2. Melgar, D., Taymaz, T., Ganas, A., Crowell, B., Öcalan, T., Kahraman, M., Tsironi, V., Yolsal-Çevikbilen, S., Valkaniotis, S., Irmak, T. S., Eken, T., Erman, C., Özkan, B., Dogan, A. H., & Altuntaş, C. (2023). Sub- and super-shear ruptures during the 2023 Mw 7.8 and Mw 7.6 earthquake doublet in SE Türkiye. Seismica, 2(3). https://doi.org/10.26443/seismica.v2i3.387
3. https://www.npr.org/2023/02/17/1157790902/turkey-earthquake-engineers-building-damage-inspectors
4. https://www.eastmojo.com/world/2023/03/16/deadly-turkey-earthquakes-put-spotlight-on-building-code-violations/
5. Taftsoglou Maria, Valkaniotis Sotiris, Karantanellis Efstratios, Goula Evmorfia, & Papathanassiou. (2023). Preliminary mapping of liquefaction phenomena triggered by the February 6 2023 M7.7 earthquake, Türkiye / Syria, based on remote sensing data. Zenodo. https://doi.org/10.5281/zenodo.7668401